\documentstyle[11pt,aas2pp4,flushrt]{article}

\newcommand{\msun}{$\hbox{M}_{\odot}$}
\newcommand{\zsun}{$\hbox{Z}_{\odot}$}
\newcommand{\PVdblt}{{\rm P}\kern 0.1em{\sc v}~$\lambda\lambda 1117, 1128$}
\newcommand{\CaIIdblt}{{\rm Ca}\kern 0.1em{\sc ii}~$\lambda\lambda 3934, 3969$}
\newcommand{\AlIIIdblt}{{\rm Al}\kern 0.1em{\sc iii}~$\lambda\lambda 1854, 1862$}
\newcommand{\CIVdblt}{{\rm C}\kern 0.1em{\sc iv}~$\lambda\lambda 1548, 1550$}
\newcommand{\MgIIdblt}{{\rm Mg}\kern 0.1em{\sc ii}~$\lambda\lambda 2796, 2803$}
\newcommand{\NVdblt}{{\rm N}\kern 0.1em{\sc v}~$\lambda\lambda 1238, 1242$}  
\newcommand{\SVIdblt}{{\rm S}\kern 0.1em{\sc vi}~$\lambda\lambda 933, 944$} 
\newcommand{\OVIdblt}{{\rm O}\kern 0.1em{\sc vi}~$\lambda\lambda 1031, 1037$} 
\newcommand{\SiIIdblt}{{\rm Si}\kern 0.1em{\sc ii}~$\lambda\lambda 1190, 1193$} 
\newcommand{\SiIVdblt}{{\rm Si}\kern 0.1em{\sc iv}~$\lambda\lambda 1393, 1402$} 
\newcommand{\PV}{\hbox{{\rm P}\kern 0.1em{\sc v}}}
\newcommand{\AlI}{\hbox{{\rm Al}\kern 0.1em{\sc i}}}
\newcommand{\AlII}{\hbox{{\rm Al}\kern 0.1em{\sc ii}}}
\newcommand{\AlIII}{{\hbox{\rm Al}\kern 0.1em{\sc iii}}}
\newcommand{\CaII}{\hbox{{\rm Ca}\kern 0.1em{\sc ii}}}
\newcommand{\CII}{\hbox{{\rm C}\kern 0.1em{\sc ii}}}
\newcommand{\CIIe}{\hbox{{\rm C$^{\ast}$}\kern 0.1em{\sc ii}}}
\newcommand{\CIII}{\hbox{{\rm C}\kern 0.1em{\sc iii}}}
\newcommand{\CIV}{\hbox{{\rm C}\kern 0.1em{\sc iv}}}
\newcommand{\CV}{\hbox{{\rm C}\kern 0.1em{\sc v}}}
\newcommand{\HI}{\hbox{{\rm H}\kern 0.1em{\sc i}}}
\newcommand{\HII}{\hbox{{\rm H}\kern 0.1em{\sc ii}}}
\newcommand{\Lya}{\hbox{{\rm Ly}\kern 0.1em$\alpha$}}
\newcommand{\Lyb}{\hbox{{\rm Ly}\kern 0.1em$\beta$}}
\newcommand{\Lyg}{\hbox{{\rm Ly}\kern 0.1em$\gamma$}}
\newcommand{\Lyd}{\hbox{{\rm Ly}\kern 0.1em$\delta$}}
\newcommand{\Lye}{\hbox{{\rm Ly}\kern 0.1em$\epsilon$}}
\newcommand{\Lyphi}{\hbox{{\rm Ly}\kern 0.1em$\phi$}}
\newcommand{\Lyfive}{\hbox{{\rm Ly}\kern 0.1em$5$}}
\newcommand{\Lysix}{\hbox{{\rm Ly}\kern 0.1em$6$}}
\newcommand{\Lyseven}{\hbox{{\rm Ly}\kern 0.1em$7$}}
\newcommand{\Lyeight}{\hbox{{\rm Ly}\kern 0.1em$8$}}
\newcommand{\Lynine}{\hbox{{\rm Ly}\kern 0.1em$9$}}
\newcommand{\Lyten}{\hbox{{\rm Ly}\kern 0.1em$10$}}
\newcommand{\Lyeleven}{\hbox{{\rm Ly}\kern 0.1em$11$}}
\newcommand{\HeI}{\hbox{{\rm He}\kern 0.1em{\sc i}}}
\newcommand{\HeII}{\hbox{{\rm He}\kern 0.1em{\sc ii}}}
\newcommand{\FeI}{\hbox{{\rm Fe}\kern 0.1em{\sc i}}}
\newcommand{\FeII}{\hbox{{\rm Fe}\kern 0.1em{\sc ii}}}
\newcommand{\FeIII}{\hbox{{\rm Fe}\kern 0.1em{\sc iii}}}
\newcommand{\MnII}{\hbox{{\rm Mn}\kern 0.1em{\sc ii}}}
\newcommand{\MgI}{\hbox{{\rm Mg}\kern 0.1em{\sc i}}}
\newcommand{\MgII}{\hbox{{\rm Mg}\kern 0.1em{\sc ii}}}
\newcommand{\MgIII}{\hbox{{\rm Mg}\kern 0.1em{\sc iii}}}
\newcommand{\NI}{\hbox{{\rm N}\kern 0.1em{\sc i}}}
\newcommand{\NII}{\hbox{{\rm N}\kern 0.1em{\sc ii}}}
\newcommand{\NIII}{\hbox{{\rm N}\kern 0.1em{\sc iii}}}
\newcommand{\NV}{\hbox{{\rm N}\kern 0.1em{\sc v}}}
\newcommand{\OVI}{\hbox{{\rm O}\kern 0.1em{\sc vi}}}
\newcommand{\OI}{\hbox{{\rm O}\kern 0.1em{\sc i}}}
\newcommand{\OII}{\hbox{[{\rm O}\kern 0.1em{\sc ii}]}}
\newcommand{\OIV}{\hbox{{\rm O}\kern 0.1em{\sc iv}]}}
\newcommand{\SI}{{\rm S}\kern 0.1em{\sc i}}
\newcommand{\SIV}{{\rm S}\kern 0.1em{\sc iv}}
\newcommand{\SVI}{{\rm S}\kern 0.1em{\sc vi}}
\newcommand{\SiI}{\hbox{{\rm Si}\kern 0.1em{\sc i}}}
\newcommand{\SiII}{\hbox{{\rm Si}\kern 0.1em{\sc ii}}}
\newcommand{\SiIII}{\hbox{{\rm Si}\kern 0.1em{\sc iii}}}
\newcommand{\SiIV}{\hbox{{\rm Si}\kern 0.1em{\sc iv}}}
\newcommand{\SII}{\hbox{{\rm S}\kern 0.1em{\sc ii}}}
\newcommand{\SIII}{\hbox{{\rm S}\kern 0.1em{\sc iii}}}
\newcommand{\NaI}{\hbox{{\rm Na}\kern 0.1em{\sc i}}}
\newcommand{\TiII}{\hbox{{\rm Ti}\kern 0.1em{\sc ii}}}
\newcommand{\kms}{\hbox{km~s$^{-1}$}}
\newcommand{\sqcm}{\hbox{cm$^{2}$}}
\newcommand{\cmsq}{\hbox{cm$^{-2}$}}

\twocolumn

\begin{document}

\title{Quasistellar Objects: Intervening Absorption Lines\altaffilmark{1}}

\author{Jane~C.~Charlton and Christopher~W.~Churchill}
\affil{The Pennsylvania State University, University Park, PA 16802}
%charlton@astro.psu.edu, cwc@astro.psu.edu

\altaffiltext{1}{Written for the Encyclopedia of Astronomy and Astrophysics
(to be published in 2000 by MacMillan and the Institute of Physics
Publishing)}

\begin{abstract}
{\small We briefly review, at a level appropriate for graduate students and
non-specialists, the field of quasar absorption lines (QALs).  Emphasis
is on the intervening absorbers.  We present the anatomy of a quasar
spectrum due to various classes of intervening absorption systems, and
a brief historical review of each absorber class (Lyman-alpha forest and
Lyman limit systems, and metal-line and damped Lyman-alpha absorbers).
We also provide several heuristic examples on how the physical properties
of both the intergalactic medium and the gaseous environments associated
with earlier epoch galaxies can be inferred from QALs.  The evolution of
these environments from z=4 are discussed.}
\end{abstract}

\section{Introduction}

Every parcel of gas along the line of sight to a distant quasar
will selectively absorb certain wavelengths of continuum light 
of the quasar due to the presence of the various chemical 
elements in the gas.
Through the analysis of these quasar absorption lines we can
study the spatial distributions, motions, chemical enrichment,
and ionization histories of gaseous structures from redshift
five until the present.
This includes the gas in galaxies of all morphological types
as well as the diffuse gas in the intergalactic medium.

\subsection{Basics of Quasar Spectra}

\begin{figure*}[th]
\figurenum{1}
\plotfiddle{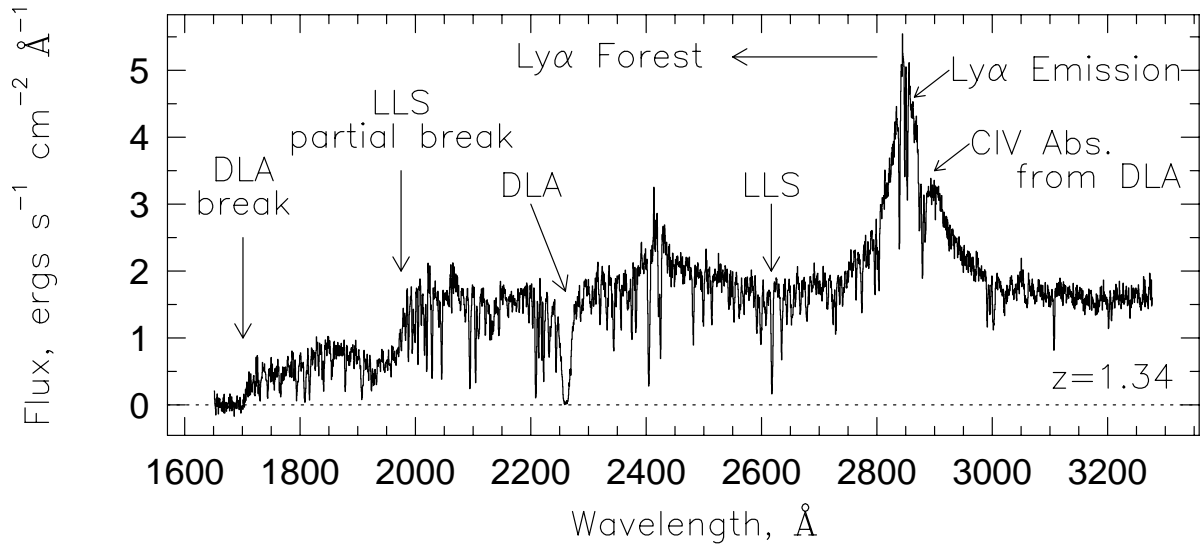}{3.0in}{0}{90.}{90.}{-270}{-40}
\vglue -0.15in
\caption{\footnotesize
Typical spectrum of a quasar, showing the quasar continuum
and emission lines, and the absorption lines produced by galaxies 
and intergalactic material that lie between the quasar and the observer.
This spectrum of the $z=1.34$ quasar PKS$0454+039$ was obtained with the
Faint Object Spectrograph on the Hubble Space Telescope. The emission
lines at $\sim 2400$~{\AA} and $\sim 2850$~{\AA} are
{\Lyb} and {\Lya}.  The {\Lya} forest, absorption
produced by various intergalactic clouds, is apparent at
wavelengths blueward of the {\Lya} emission line.
The two strongest absorbers, due to galaxies, are a damped
{\Lya} absorber at $z=0.86$ and a Lyman limit system
at $z=1.15$.  The former produces a Lyman limit break at 
$\sim1700$~{\AA} and the latter a partial Lyman limit break
at $\sim1950$~{\AA} since the neutral Hydrogen column density
is not large enough for it to absorb all ionizing photons.
Many absorption lines are produced by the DLA at $z=0.86$
({\CIVdblt}, for example, is redshifted onto the red wing of the
quasar's {\Lya} emission line).}
\label{fig:fig1.eps}
\end{figure*}

Figure~1 illustrates many of the common features of a quasar
spectrum.  The relatively flat quasar continuum and broad emission
features are produced by the quasar itself (near the black hole and
its accretion disk).
In some cases, gas near the quasar central engine also
produces ``intrinsic'' absorption lines, 
most notably {\Lya}, and relatively high ionization 
metal transitions such as {\CIV}, {\NV}, and {\OVI}.
These intrinsic absorption lines can be broad [thousands
or even tens of thousands of {\kms} in which case the quasar 
is called a broad absorption line (BAL) QSO], or narrow 
(tens to hundreds of {\kms}).
However, the vast majority of absorption lines in a typical
quasar spectrum are ``intervening'', produced by
gas unrelated to the quasar that is located along the
line of sight between the quasar and the Earth.

A structure along the line of sight to the quasar can
be described by its neutral Hydrogen column density, $N({\HI})$,
the number of atoms per {\sqcm}.  $N({\HI})$ is given by the
product of the density of the material and the pathlength
along the line of sight through the gas.  Each structure 
will produce an absorption line in the quasar spectrum at a 
wavelength of $\lambda_{obs} = \lambda_{rest} (1+z_{abs})$, where 
$z_{abs}$ is the redshift of the absorbing gas and 
$\lambda_{rest}=1215.67$~{\AA} is the rest wavelength of the 
{\Lya} transition.
Since $z_{abs} < z_{QSO}$, the redshift of the quasar, these
{\Lya} absorption lines form a ``forest'' at wavelengths
blueward of the {\Lya} emission.
The region redward of the {\Lya} emission will be
populated only by absorption through other chemical transitions
with longer $\lambda_{rest}$.
Historically, absorption systems with $N({\HI}) < 10^{17.2}$~{\cmsq}
have been called {\Lya} forest lines, those with
$10^{17.2} < N({\HI}) < 10^{20.3}$~{\cmsq} are Lyman limit systems, 
and those with $N({\HI}) > 10^{20.3}$~{\cmsq} are damped {\Lya} systems.
The number of systems per unit redshift increases dramatically with
decreasing column density, as illustrated in the schematic diagram
in Figure~2.
Lyman limit systems are defined by a sharp break in the spectrum
due to absorption of photons capable of ionizing {\HI}, i.e.
those with energies greater than $13.6~{\rm eV}$.
The optical depth, $\tau$, of the break is given by the product 
$N({\HI}) \sigma$, where the cross section for ionization of Hydrogen,
$\sigma = 6.3 \times 10^{-18} (E_{\gamma}/13.6~{\rm eV})^{-3}~{\sqcm}$,
(and the flux is reduced by the factor $e^{-\tau}$).
The energy dependence of $\sigma$ leads to a recovery of the 
Lyman limit break at higher energies (shorter wavelengths), unless 
$N({\HI}) \gg 10^{17.2}$~{\cmsq} (see Figure~1).

The curve of growth describes the relationship between the
equivalent width of an absorption line, $W$, (the integral of
the normalized profile) and its column density, $N$.
Figure~3 shows that for small $N({\HI})$ the number of absorbed
photons, and therefore the flux removed, increases in direct 
proportion to the number of atoms.  
This is called the linear part of the curve of growth.
As $N$ is increased the line saturates so that photons are only 
absorbed in the wings of the lines; in this regime the equivalent 
width is sensitive to the amount of line broadening
(characterized by the Doppler parameter $b$), but does
not depend very strongly on $N({\HI})$.
This is the flat part of the curve of growth.
Finally, at $N({\HI}) > 10^{20.3}$~{\cmsq}, there are enough atoms
that the damping wings of the line become populated and the 
equivalent width increases as the square root of $N({\HI})$,
and is no longer sensitive to $b$.

\begin{figure}[th]
\figurenum{2}
\plotone{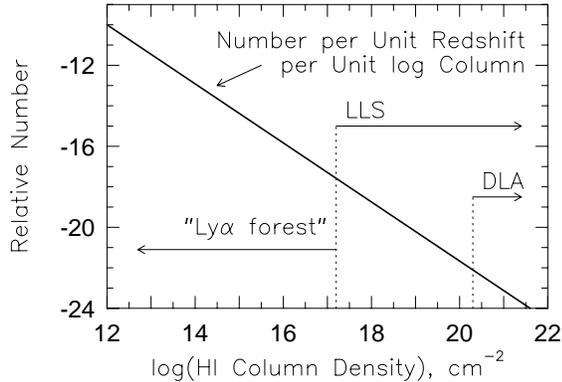}
\caption{\footnotesize
The column density distribution of {\Lya} clouds, $f(N({\HI})$,
roughly follows a power law over ten orders of magnitude;
there are many more weak lines than strong lines.
The column density regions for the three categories of systems
are shown: {\Lya} forest, Lyman limit, and damped {\Lya}.
The term ``{\Lya} forest'' has at times been used to
refer to metal--free Hydrogen clouds, perhaps those with
$N({\HI}) < 10^{16}$~{\cmsq}, but now metals have been
found associated with weaker systems down to the detection
limit.}
\label{fig:fig2}
\end{figure}

In addition to the {\Lya} ($1s \rightarrow 2p$)
and higher order ($1s \rightarrow np$) Lyman series
lines, quasar spectra also show absorption due to different
ionization states of the various species of metals.
Figure~1 illustrates that the damped {\Lya} system at 
$z=0.86$ that is responsible for the
{\Lya} absorption line at $\lambda_{obs} = 2260$~{\AA} and
a Lyman limit break at $\lambda_{obs} = 1700$~{\AA}
also produces absorption at $\lambda_{obs} = 2870$~{\AA} due 
to the presence of {\CIV} in the absorbing gas at that same redshift.
Like many of the strongest metal lines seen in quasar
spectra, {\CIV} is a resonant doublet transition due to
transitions from $^2S_{1/2}$ energy levels to the $^2P_{1/2}$ 
and to the $^2P_{3/2}$ energy levels.
(The left superscript ``2'' represents the number of orientations of
the electron spin, the letter $S$ or $P$ represents the total orbital angular
momentum, $L$, and the right subscript represents the total
angular momentum, $J$.)
Doublet transitions are easy to identify.  The
dichotomy between rest wavelength and redshift is 
resolved because the observed wavelength 
separation of the doublet members increases as $1+z$.

Table~1 lists some of the metal lines that are commonly
detected for intervening absorption systems.
Many of these are only strong enough to be observable
for quasar lines of sight that pass through the higher
$N({\HI})$ regions of galaxies.

\begin{center}
\vglue 0.3in
\begin{tabular*}{2.5in}[]{lr}
\multicolumn{2}{c}{Table 1: Common Transitions} \\
\hline\hline
Transition\phantom{...........} & $\lambda_{\rm rest}$~[{\AA}] \\
\hline
{\rm LL} \dotfill & $\sim912$ \\
{\Lyg}\dotfill & 972.537  \\
{\Lyb}\dotfill & 1025.722  \\
{\Lya}\dotfill & 1215.670  \\
{\SiIV}~1393\dotfill & 1393.755 \\
{\SiIV}~1402\dotfill & 1402.770 \\
{\CIV}~1548\dotfill & 1548.195  \\
{\CIV}~1550\dotfill & 1550.770  \\
{\FeII}~2382\dotfill & 2382.765  \\
{\FeII}~2600\dotfill & 2600.173  \\
{\MgII}~2796\dotfill & 2796.352  \\
{\MgII}~2803\dotfill & 2803.531  \\ \hline
\end{tabular*}
%\begin{tabular*}{2.5in}[]{lcl}
%\multicolumn{3}{c}{Table 1: Atomic Transitions} \\
%\hline\hline
%Transition & $\lambda_{\rm rest}$~[{\AA}] & $f_{\rm osc}$ \\
%\hline
%{\rm LL} & $\sim912$ & \\
%{\Lyg} & 972.537 & 0.029 \\
%{\Lyb} & 1025.722 & 0.079 \\
%{\Lya} & 1215.670 & 0.416 \\
%{\CIV}~1548 & 1548.195 & 0.191 \\
%{\CIV}~1550 & 1550.770 & 0.095 \\
%{\FeII}~2382 & 2382.765 & 0.301 \\
%{\FeII}~2600 & 2600.173 & 0.224 \\
%{\MgII}~2796 & 2796.352 & 0.612 \\
%{\MgII}~2803 & 2803.531 & 0.305 \\ \hline
%\end{tabular*}
\end{center}

\begin{figure*}[pht]
\figurenum{3}
\plotfiddle{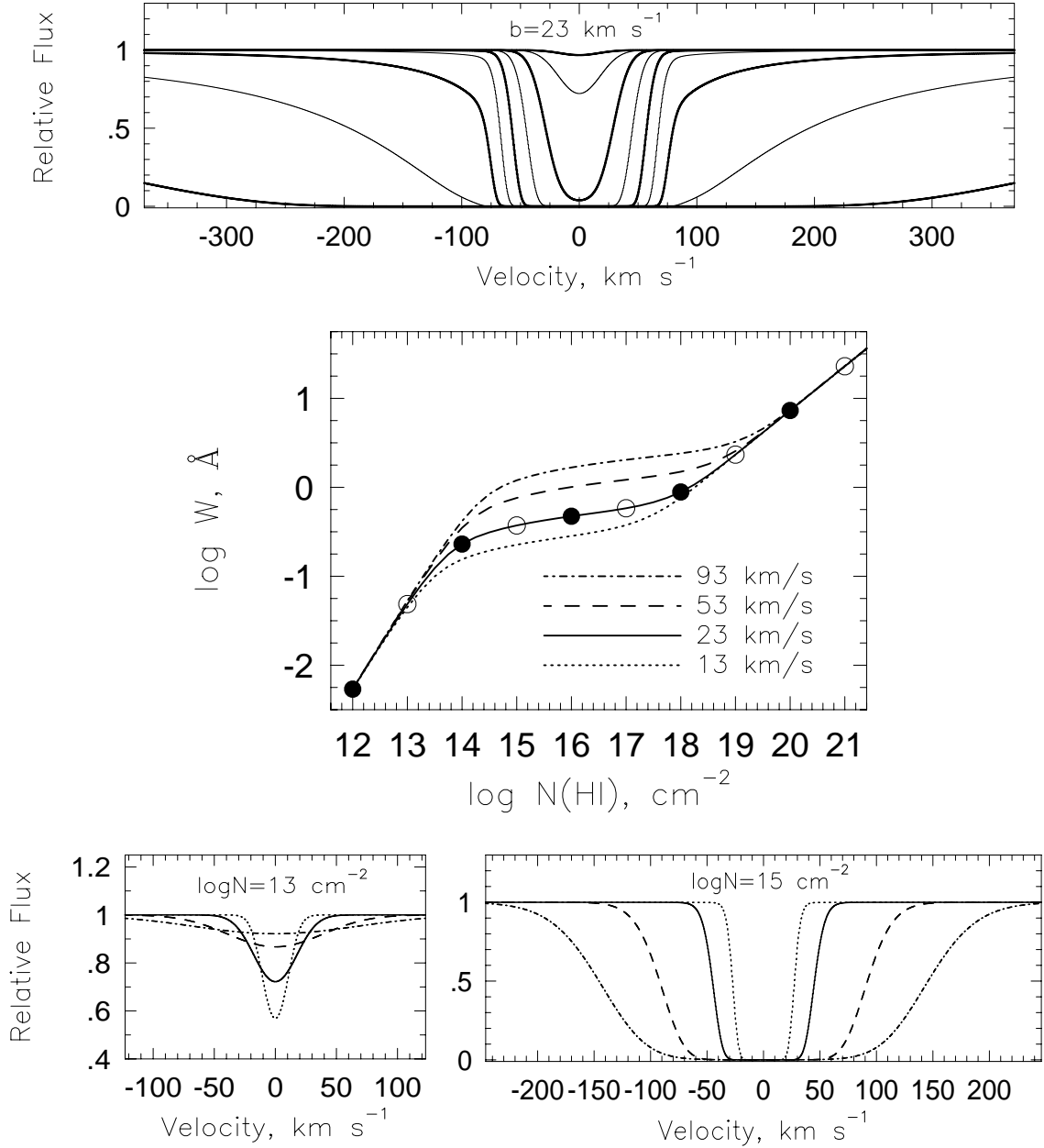}{5.5in}{0}{90.}{90.}{-243}{-50}
\vglue 0.56in
\caption{\footnotesize
Illustration of the different regimes of the curve of
growth.  
The middle panel shows the curve of growth for the {\Lya}
transition, relating the equivalent width, $W$, of the absorption
profile to the column density, $N({\HI})$.  The different curves
represent four different values of the Doppler parameter:
$b=13$, $23$, $53$, and $93$~{\kms}.  
The upper panel shows absorption profiles with
Doppler parameter $b=23$~{\kms} for the series of neutral hydrogen 
column densities $N({\HI}) = 10^{12}$ -- $10^{20}$~{\cmsq}.
The thick (thin) curves correspond to the filled (open) points 
on the $b=23$~{\kms} curve of growth (middle panel), starting at 
$N({\HI}) = 10^{12}$~{\cmsq}.
For $N({\HI}) < 10^{13}$~{\cmsq}, 
known as the linear part of the curve of growth, the equivalent width
does not depend on $b$.  The lower left panel shows that, at fixed
$N({\HI})$, the depth of the profile is smaller for large $b$,
such that the equivalent width remains constant.  On the flat
part of the curve of growth, profiles are saturated and
the equivalent width increases with $b$ for constant $N({\HI})$.
For $N({\HI}) > 10^{20}$~{\cmsq}, the profile develops damping
wings, which dominate the equivalent width.}
\label{fig:fig3}
\end{figure*}

\section{History, Surveys, and Revolutionary Progress in the 1990's}

The history of quasar absorption lines began within a couple of
years of the identification of the first quasar in 1963.
In 1965, Gunn and Peterson considered the detection of flux
blueward of the {\Lya} emission line in the quasar
3C~9, observed by Schmidt, and derived a limit on the
amount of neutral Hydrogen that could be present in intergalactic
space.
In that same year, Bahcall and Salpeter predicted
that intervening material should produce observable discrete
absorption features in quasar spectra.
Such features were detected in 1967 in the quasar PKS~$0237-23$ by
Greenstein and Schmidt, and in 1968 in PHL~$938$ by Burbidge,
Lynds, and Stockton.
By 1969 many intervening systems had been discovered, and
Bahcall and Spitzer proposed that most with metals were produced
by the halos of normal galaxies.
As more data accumulated, the sheer number of {\Lya} forest
lines strongly supported the idea that galactic and intergalactic
gas, and not only material intrinsic to the quasar, is the
source of most quasar absorption lines.  
%In 1969, Bahcall and Peebles developed a statistical test to study
%the clustering properties of absorption lines,
%and in 1980 Sargent and collaborators applied this test to
%find that the {\Lya} forest lines are randomly distributed 
%along quasar lines of sight, and not clustered like galaxies.

In the 1980's many more quasar spectra were obtained and
many large statistical surveys of the different classes of
absorption line systems were published.
The emphasis was to characterize the number of lines per unit 
redshift, $dN/dz$, stronger than some specified equivalent width 
limit.
With 4m--class telescopes [equipped with charge coupled device
(CCD) detectors] it was possible to conduct surveys with a
spectral resolution of $R\sim1000$.
The spectral resolution is defined as 
$R = \lambda/\Delta \lambda = c/\Delta v$, so that $R=1000$
corresponds to $300$~{\kms} or $5$~{\AA} at $\lambda = 5000$~{\AA}.
Separate surveys were conducted for {\Lya} lines, {\MgII} doublets,
{\CIV} doublets, and also for Lyman limit breaks, all as
a function of redshift.
The {\Lya} line is observable in the optical part of the
spectrum for $z>2.2$, {\MgII} for $0.4 < z < 2.2$, {\CIV} for
$1.7 < z < 5.0$, and the Lyman limit break for $z > 3$.
However, a break is also easily identified in lower resolution
space--based UV spectra, which extended Lyman limit surveys to 
lower redshift.

In order to consider the cross section of the sky covered
by the different populations, it can be assumed that absorption 
will be observed for all lines of sight within some radius of 
every luminous galaxy ($>0.05 L_K^*$). ($L_K^*$ represents
the Schechter luminosity, i.e. the transition between the 
exponential and the power law forms of the luminosity function,
and corresponds to a $K$--band absolute magnitude of $M_K = -25$).
To explain the observed $dN/dz$ at $z \sim 1.5$, this radius would be
$70$~kpc for strong {\CIV} (detection sensitivity $0.4$~{\AA}), 
and $40$~kpc for strong {\MgII} (detection sensitivity $0.3$~{\AA})
and also for Lyman limit systems, implying that the latter two 
populations are in fact produced in the same gas.
The higher $N({\HI})$ damped {\Lya} absorbers would be
produced within $15$~kpc of the center of each galaxy, while
the {\Lya} forest lines would require a considerably
larger region, hundreds of kpcs around each galaxy to produce
a cross section consistent with the observed number of weak lines.

Up until the 1990's, the focus of quasar absorption line
work was to separately consider the properties of the individual
classes of absorbers (eg.\ {\Lya} forest or {\MgII} absorbers).
In the 1990's, however, three different observational advances 
led to recognition of the direct connections between the different 
classes of quasar absorption lines, and of direct associations 
with the population of galaxies:

1. Deep images of quasar fields could be obtained, and redshifts
of the galaxies in the field could be determined from low resolution
spectra.
Steidel found that whenever {\MgII} absorption 
with $W_r({\MgII}) > 0.3$~{\AA} is observed, a luminous galaxy
($L_K > 0.06 L_K^*$) is found within an impact parameter of 
$38 {\rm h}^{-1} (L/L_K^*)^{-0.15}~{\rm kpc}$ with a redshift 
coincident with that determined from the absorption lines.
Also, it is rare to find a galaxy within this impact parameter
that does not produce {\MgII} absorption.
There appears to be a one--to--one correspondence between 
strong {\MgII} absorption and luminous galaxies.
The {\MgII} absorbing galaxies span a range of morphological 
types.

2. The High Resolution Spectrograph on the Keck I
$10$-meter telescope made it possible to obtain quasar spectra at a
resolution of $R = 45,000$, which corresponds to $\sim 6$~{\kms}.
The previous surveys with resolution of order hundreds of
{\kms} identified absorption due to entire galaxies and their
environments.
With $6$~{\kms} resolution it became possible to resolve
structure within a galaxy: the clouds in its halo, the
interstellar medium of its disk, and the satellites and
infalling gas clouds in its environment.
Figure~4 is a dramatic illustration of this contrast for
the {\MgII} absorber at $z=0.93$ toward the quasar PG~$1206+459$.

\begin{figure*}[pt]
\figurenum{4}
\plotfiddle{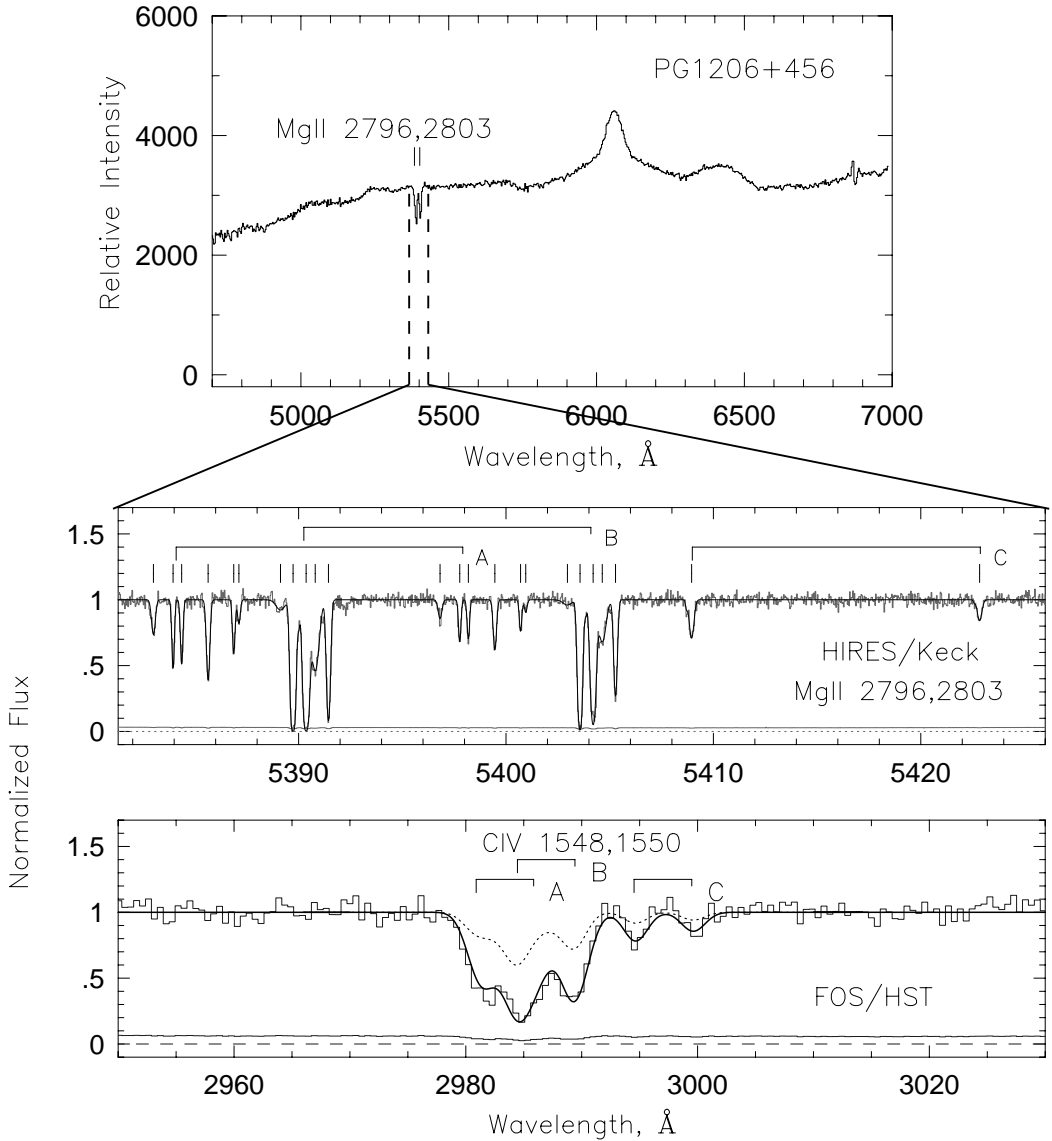}{9.4in}{0}{80.}{80.}{-210}{220}
\vglue -3.6in
\caption{\footnotesize
Dramatic demonstration of gains due to high resolution
spectroscopy of the {\MgII} doublet.  The top panel is a $R=3000$
spectrum of PG$1206+459$.  The doublet that is apparent at
an observed wavelength of $\sim 5400$~{\AA} is due to {\MgII}
absorption from a system at $z=0.927$.  The middle panel shows
the remarkable kinematic structure that is revealed at the
resolution ($R=45,000$) of the Keck/HIRES spectrograph of
the same quasar.  The $2796$~{\AA}
transition is resolved into multiple components ($5583$--$5592$~{\AA}),
which also appear in the $2803$~{\AA}
transition ($5396$--$5406$~{\AA}).  This system can be separated
in two ``clusters'' of clouds, labeled ``A'' and ``B''.
Another weaker {\MgII} doublet is observed at $5409$ and 
$5423$~{\AA}, from a system at $z=0.934$~{\AA}, labeled with a 
``C''.
The solid line through these complex {\MgII} profiles is the
result of multiple Voigt profile fitting, with a cloud centered 
on each of the ticks drawn above the spectrum.
The lower panel shows the {\CIV} doublets associated
with the same three systems, observed with the Faint Object
Spectrograph on HST, but at much lower resolution ($R=1300$).
The {\CIV} is in three different concentrations around the
three systems ``A'', ``B'', and ``C''.  The {\CIV}$\lambda 1550$
transition from system A is blended with the 
{\CIV}$\lambda 1548$ transition from system B.
The {\CIV} equivalent width is too large for this
absorption to be produced by the same phase of gas that
produces the {\MgII} cloud absorption.
The maximum absorption that can arise in the {\MgII}
phase is given by the dotted line; a plausible
model with a kinematically broader {\CIV} phase yields
the solid curve.}
\label{fig:fig4}
\end{figure*}

3. The Faint Object Spectrograph (FOS) 
%and Goddard High Resolution Spectrograph (GHRS) 
on the Hubble Space Telescope provided resolution $R\sim1000$
in the UV, from $1400$--$3300$~{\AA}.
Observations of {\Lya} forest clouds could be extended from 
$z=2.2$ down to the present epoch.
Furthermore, absorption from a given galaxy could be observed in 
numerous transitions; if {\MgII} was observed in the optical,
the Lyman series and {\CIV} could be studied in the UV
(see Figure~4).
With information on transitions with a range of ionization
states, consideration of the degree of ionization (related
to the gas density and the intensity and shape of the 
ionizing radiation field) and the multiple phase structure
of galactic gas became possible.

No longer is analysis of absorption lines in quasar spectra
an esoteric subject.  It has developed into a powerful tool to 
be used in the study of galaxy evolution (eg.\ similar to 
imaging the stellar components of the galaxies).
At least in principle, quasar spectra can be used for an
unbiased study of the gaseous environments of galaxies from
the present back to the highest redshifts at which quasars
are observed.
Gas structures smaller than $1$~{\msun} can be detected if they
are intercepted by the quasar line of sight, irrespective of
whether they emit light.
Through the tool of quasar absorption lines, proto--galactic 
structures and low surface brightness galaxies can be studied
as well as high luminosity galaxies.

\section{Developing Physical Intuition}

With high resolution spectra of quasars, it is possible to
consider the physical conditions of the gaseous structures that
produce absorption.
However, it is challenging to separate the various effects that
``shape'' the spectral features in the different chemical
transitions.
The absorption profiles observed for the different chemical
transitions are determined by a combination of the spatial
distribution of material along the line of sight, its
bulk kinematics, temperature, metallicity, and abundance
pattern.
The ionization structure is influenced by gas densities and
by the UV radiation field, which is a combination of the 
extragalactic background
radiation due to the accumulated effect of quasars and
stellar photons escaped from galaxies (and corrected for
absorption by the intergalactic medium).

The shape of an absorption line can be modeled with a
Voigt profile, which is a combination
of the natural, quantum mechanical Lorentzian broadening
and the Gaussian broadening caused by the thermal and
turbulent motions in the gas.
Several Voigt profiles can be blended together to form an
overall complex absorption feature (see Figure~4).
The ``width'' of a single Voigt profile is characterized by the
Doppler parameter, $b$ (expressed in velocity units and related 
to the Gaussian $\sigma$ by $b= 2^{1/2} \sigma$).
Physically, the Doppler parameter is the sum of thermal and turbulent
components, $b^2_{tot} = 2kT/m + b^2_{turb}$, where
$T$ is the temperature of the gas, and $m$ is the mass of an atom.

\subsection{Kinematic Models}

\begin{figure*}[pth]
\figurenum{5}
\plotfiddle{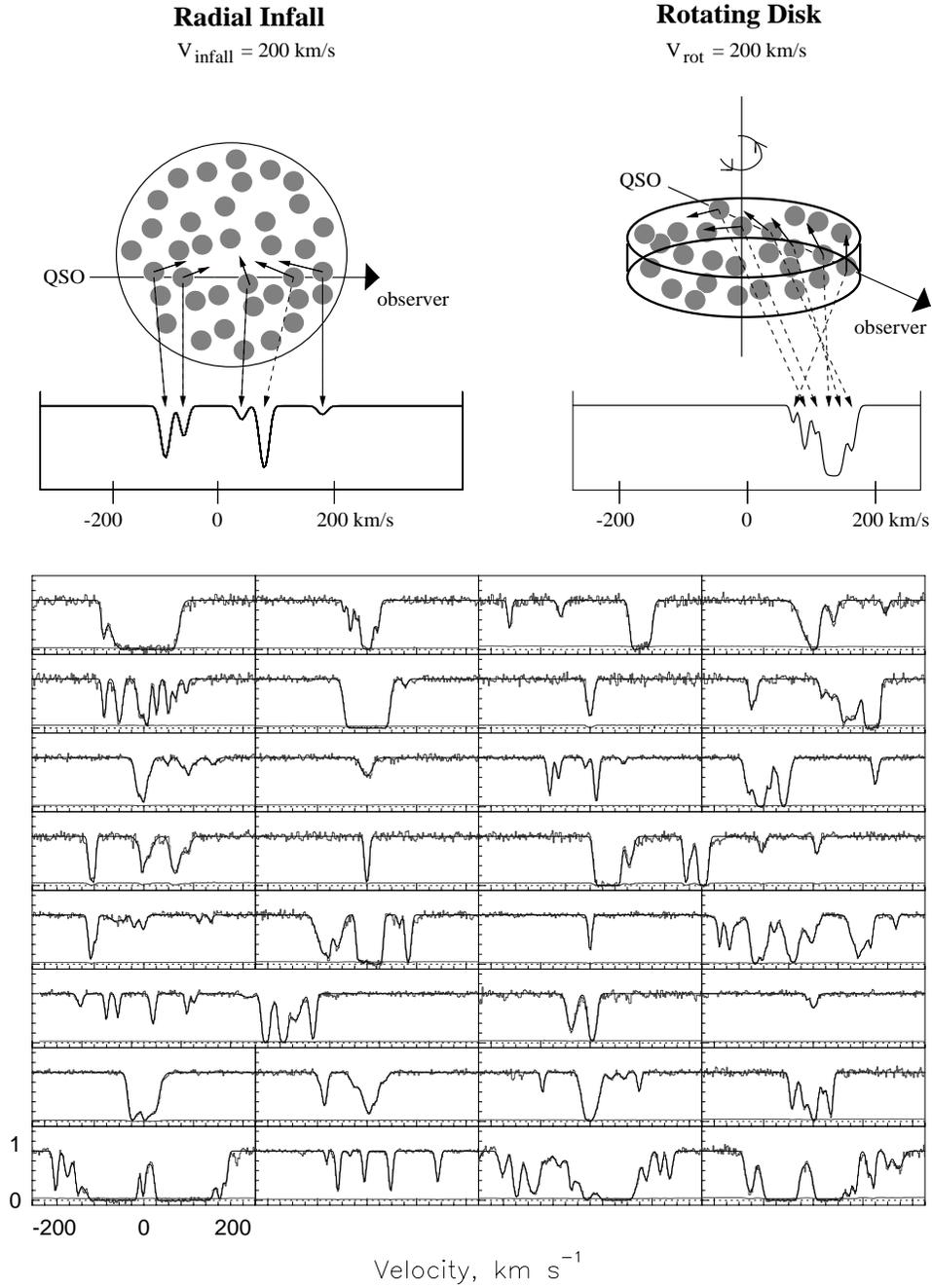}{9.4in}{0}{80.}{80.}{-190}{205}
\vglue -3.0in
\caption{\footnotesize
Illustrations of two simple kinematic models are
shown in the top panel.  To the left, the model is radial
infall of clouds to the center of a sphere, with constant
velocity.  The line of sight passes through five clouds,
which leads to five different absorption features (for a
single transition) in the quasar spectrum.  Two of the
features are blueshifted relative to the standard of rest
of the absorbing galaxy, and the other three are redshifted.
The absorption features from a radial infall model can
be spread over a velocity of $100$--$200$~{\kms}, typical
of the velocity dispersion of a galaxy halo.
To the right, a rotating disk model is illustrated.  In this
case all the ``clouds'' along the line of site have a
component of motion that is redshifted, and they tend to
be clustered together in velocity space, with typical
spread of $20$-$60$~{\kms}.
The lower panel shows a sample of $0.4 < z < 1.4$ {\MgII}
absorption profiles observed with the Keck/HIRES spectrograph
at $R=45,000$, corresponding to a resolution of $\sim 6$~{\kms}.
The solid lines through these data are Voigt profile
fits and the ticks drawn above the spectrum represent the cloud
velocities.
Some of these profiles are consistent with the kinematics of
a rotating disk, and others with radial infall kinematics.
However, to explain the full ensemble of profiles a model
combining these two basic types of kinematics is needed.}
\label{fig:fig5}
\end{figure*}

Two of the simplest types of organized kinematics in
galaxies are illustrated in Figure~5: clouds distributed in a 
rotating disk, and radial infall of clouds in a spherical 
distribution.  Here, {\MgII} absorbers are used as an 
example, but the same kinematic arguments would apply
to other transitions.
For radial infall, clouds can be distributed over the
range of velocities, with a tendency for a ``double peak''
from material that is redshifted and blueshifted
but with a considerable amount of variation if there
are typically several discrete clouds along the line
of sight.
A rotating disk with a vertical velocity dispersion
characteristic of a spiral galaxy disk ($10$--$20$~{\kms})
will have clouds superimposed in velocity space, and
an overall kinematic spread of tens of {\kms}.
Strong {\MgII} absorption has been found to arise along
nearly all lines of within $\sim40$~kpc of normal galaxies
(i.e. the covering factor is nearly unity within that radius).
The large variety of kinematics evident in {\MgII} 
absorption profiles is, in fact, consistent with a
superposition of disk and radial infall (halo) motions,
and not with just one or the other.
In addition to these simple, toy models, insights
can be gleaned by passing lines of sight through the
structures in cosmological N--body/hydrodynamic
simulations.  In a few studies, metals have been added
uniformly throughout the simulation box and photoionization 
models used to predict the absorption expected from different
structures.
This is especially important for establishing the
kinematics that would be observed from the process
of structure formation at high redshifts.

\subsection{Photoionization Models}

Consider a cloud of material, modeled by a plane parallel 
slab with a certain total column density of Hydrogen, 
$N(H) = N({\HI}) + N({\HII})$, and with a constant
total number density $n_H = n({\HI}) + n({\HII})$ along 
the line of sight.
The cloud is also characterized by its metallicity,
$Z$, which is the ratio of Fe/H expressed
relative to the solar value, {\zsun}, and
by an abundance pattern (the abundance ratios
of all other elements to Fe).
The degree of ionization in the gas depends upon the
intensity and shape of the spectrum of ionizing
radiation.
The intensity is characterized by the ionization
parameter, $U = n_{\gamma}/n_H$, which is the 
ratio of the number density of photons at the
Lyman edge to the number density of Hydrogen
($n_H = n_e$, where $n_e$ is the total number density 
of electrons).
The larger the value of $U$, the more ionized the gas.
Collisional ionization can also be an important
process for some absorption systems with gas
at high temperatures (hundreds of thousands of degrees). 
Photoionization equilibrium models typically yield
temperatures of tens of thousands of degrees.

\begin{figure}[pb]
\figurenum{6}
\plotone{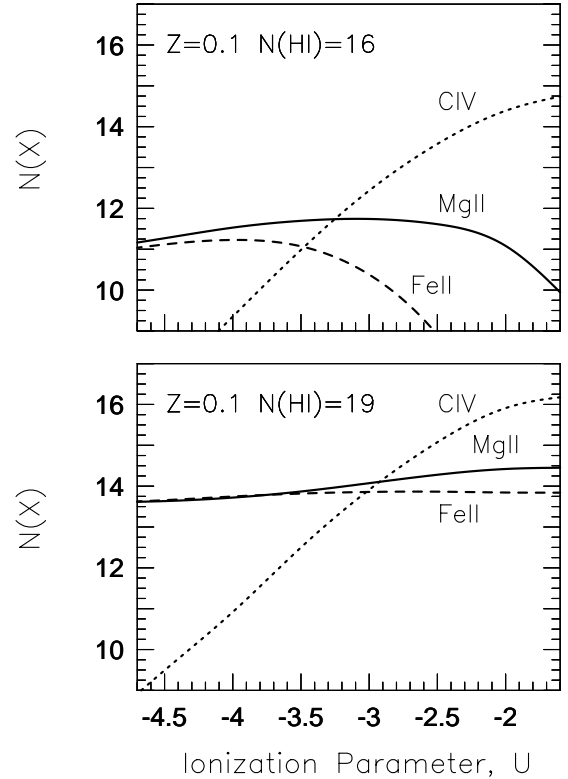}
\caption{\footnotesize
Photoionization model predictions
of the column densities of {\MgII}, {\FeII}, and {\CIV} as a
function of the ionization parameter (the ratio of ionizing
photons to the electron number density in the gas).
The spectrum incident on the cloud, represented by a constant
density slab, is the ``Haardt--Madau'' spectrum (attenuated
spectrum due to integrated effect of quasars and young galaxies).
The predicted column densities are presented in two
series of models with $N({\HI}) = 10^{16}$~{\cmsq} and with
$N({\HI}) = 10^{19}$~{\cmsq}, the optically thin and optically
thick cases.  
For both, the metallicity is fixed at $10$\% of the
solar value.
For the optically thin case, the column densities scale with 
metallicity, i.e. the ratios remain constant, but
for the optically thick case the situation is more complex.}
\label{fig:fig6}
\end{figure}

Once the metallicity, abundance pattern, ionization
parameter, and spectral shape are specified the
equations of radiative transfer can be solved 
to find the column densities of all the different
ionization states of various chemical elements.
Figure~6 illustrates, for
$N({\HI}) = 10^{16}$ and $10^{19}$~{\cmsq}, the
dependence of column densities of various transitions 
on the ionization parameter, $U$.
For optically thin gas [$N({\HI}) < 10^{17.2}$~{\cmsq}],
the column density ratios of the various metal
transitions are not dependent on the overall metallicity, 
i.e.\ the curves shift vertically in proportion to $Z$.
For optically thick gas, ionization structure
develops, with an outer ionized layer around a
neutral core, and there is no simple scaling
relation with metallicity.

In practice, if we assume that a cloud has a simple,
single phase structure, the ratios of the column
densities can be used to infer the ionization parameter,
which relates to the density of the gas.
However, the abundance pattern can differ from the solar
abundance pattern because of differing degrees of
depletion onto dust, or because of different processing 
histories.
Most of the so--called $\alpha$ particle nuclei (such as
Mg and Si) are synthesized primarily by
Type II supernovae during the early history
of a galaxy when most massive stars form and quickly
evolve to reach their end states.
On the other hand, the Fe--group elements are primarily
produced by Type Ia supernovae, and therefore build up
over a longer timescale.
In the basic picture of galaxy evolution, the halo stars
are formed early, have been enriched only by Type II
supernova, and therefore are $\alpha$--element enhanced.
Younger disk stars have incorporated also the Type Ia
processed material and therefore have relatively larger
Fe--group abundances.
Ideally, several different ionization states of
the same chemical element are observed so that
there is no ambiguity between the ionization
parameter and the abundance pattern, but this
has generally not yet been possible because of limited 
wavelength coverage at high resolution.

\begin{figure}[bth]
\figurenum{7}
\plotone{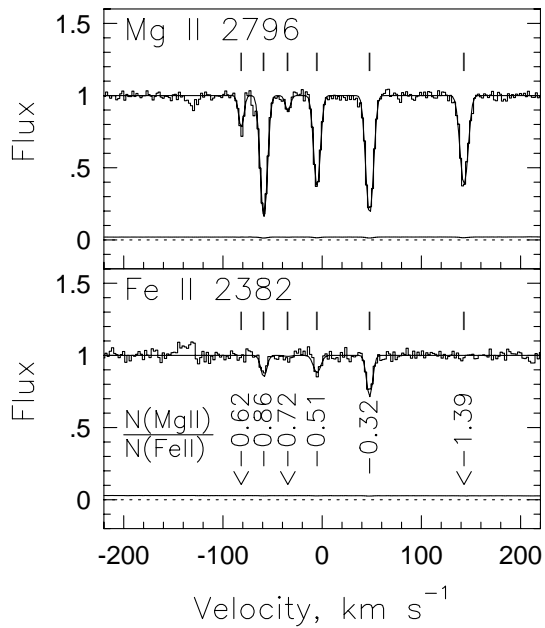}
\caption{\footnotesize
 HIRES/Keck {\FeII} and {\MgII} absorption
profiles for the $z=1.325$ system in the spectrum of the
quasar Q$0117+213$.  The six clouds in this system show a
range of more than an order of magnitude in $N({\FeII})/N({\MgII})$,
given below each cloud in the lower panel.
These variations could be due to cloud to cloud variations of
ionization parameter (density) or of abundance pattern within the
system.}
\label{fig:fig7}
\end{figure}

\begin{figure}[pth]
\figurenum{8}
\plotone{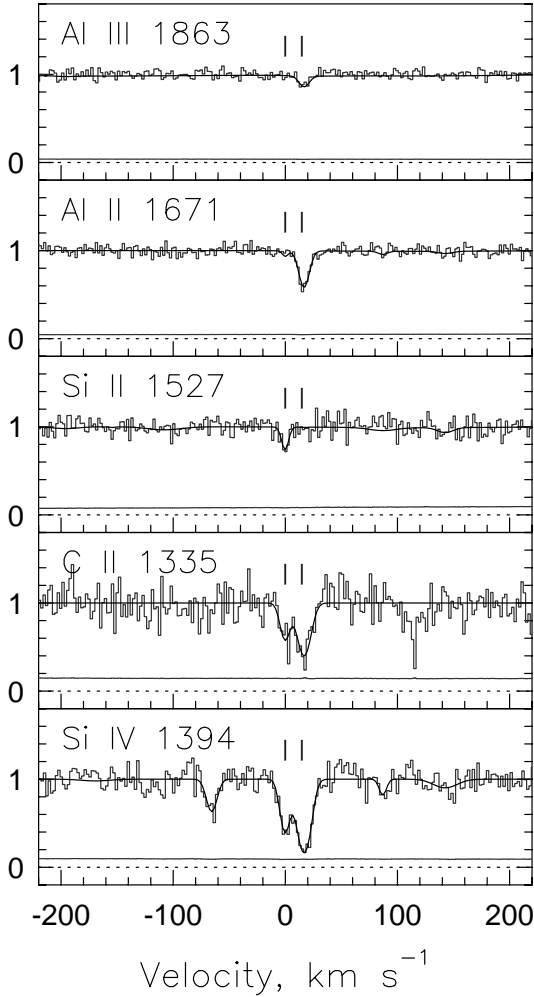}
\caption{\footnotesize
An unusual Aluminum--rich cloud is apparent in the $z=1.93$ 
system toward the quasar Q$1222+228$, and it is close in velocity space
to a normal (relative to Galactic clouds) cloud which has detected
{\SiII}.  Note the different kinematic structure in the higher
ionization transitions.  The excess of {\AlII} and {\AlIII} in the
cloud at $v=9$~{\kms} is best explained by an abundance pattern
variation, since {\SiII} and {\AlII} are transitions with very
similar ionization states.}
\label{fig:fig8}
\end{figure}

Examples of the variation of column density ratios
with velocity in two absorption systems are shown in
Figures~\ref{fig:fig7} and \ref{fig:fig8}.
In Figure 7, $N({\FeII})/N({\MgII})$ varies by
an order of magnitude over the four components in
the $z=1.325$ system toward the quasar Q$0117+213$.
This represents a variation of an order of magnitude
in the ionization parameter ($10^{-4} < U < 10^{-3}$),
or an order of magnitude variation in the abundance pattern.
Figure~8 is a very unusual system with
two clouds separated by only $20$~{\kms} in velocity,
one of which has a Silicon to Aluminum ratio similar to the 
Milky Way ISM, and the other which requires a significant 
enhancement of Aluminum.

\section{Multiphase Conditions}

The gaseous component of the the Milky Way and nearby
galaxies have phase structure (i.e.\ spatial locations with 
different densities and/or temperatures).  Examples are the
disk/halo interface (Galactic coronae) and the cold,
warm, and hot phases of the interstellar medium.
From photoionization models, it is not usually possible to
generate absorption that is simultaneously consistent with 
all observed chemical transitions for a given system.
For example, in single cloud {\MgII} systems, Figure~6
(with $N({\HI}) = 10^{16}$~{\cmsq}) shows that if {\FeII} is 
detected at a similar column density to {\MgII}, the ionization 
parameter must be small, and $W_r({\CIV})$ cannot be large.
Many systems have {\CIV} absorption which exceeds
this limit and requires a higher ionization
(lower density) phase; generally, this phase must 
have structure over a large velocity range
(a large ``effective'' Doppler parameter).
The $z=0.93$ system toward the quasar PG~1206+459
is another case that requires multiphase structure.
The observed {\CIV} profile in Figure~4 is much too strong 
for this absorption to arise in the same clouds that
produce the {\MgII}, even if their ionization parameters
are pushed to the largest values consistent with the
data.

\section{Statistics, Evolution, and Interpretation}

Future quasar absorption line studies will combine insights 
gained from detailed analyses of individual systems with 
conclusions drawn from the large statistical samples assembled 
over cosmic time.
Evolution of the ensemble of absorption profiles generated by
the universal collective of intervening structures is a result
of the combined effects of numerous processes.
These include growth of structure, star formation, morphological
evolution of galaxies, galaxy mergers, and changes in the 
extragalactic background radiation.
Here, we summarize the best present statistical data and likely 
interpretations for the different classes of absorbers.
The number of lines per unit redshift for various
populations of absorbers is represented by a power law
$dN/dz \propto (1+z)^{\gamma}$.  For a universe with only
the cosmological evolution due to expansion, $\gamma = 1.0$
for deceleration parameter $q_0 = 0$ and $\gamma = 0.5$ for
$q_0 = 0.5$.

\subsection{{\Lya} Forest}

\begin{figure*}[th]
\figurenum{9}
\plotfiddle{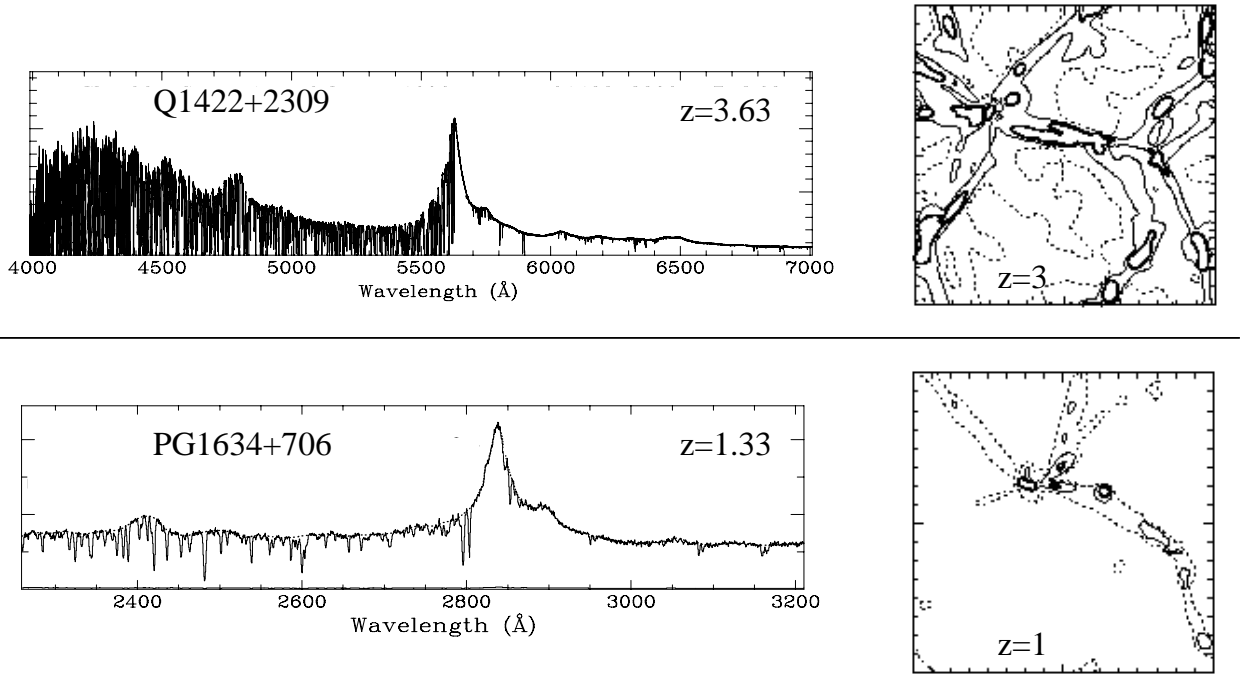}{4.0in}{270}{70.}{70.}{-240}{300}
\vglue -0.5in
\caption{\footnotesize
Illustration of structure evolution of intergalactic gas from
high to low redshift.
The upper spectrum of a $z=3.6$ quasar is a Keck/HIRES observation,
while the lower spectrum is a FOS/HST observations of a $z=1.3$ quasar.
Higher redshift quasars show a much thicker forest of {\Lya}
lines.  Slices through N--body/hydrodynamic simulation results
at the two epochs $z=3$ and $z=1$ are shown in the right--hand panels.
Three contour levels are shown:
$10^{11}$~{\cmsq} (dotted lines), $10^{12}$~{\cmsq} (solid lines) and
$10^{13}$~{\cmsq} (thick solid lines).
Evolution proceeds so that the voids become more empty so that even
the low column density material is found in filamentary structures
at low redshifts.}
\label{fig:fig9}
\end{figure*}

The {\Lya} forest evolves away dramatically from high to 
low redshift, as is strikingly clear from the spectra of
$z\sim3$ and $z\sim1$ quasars in Figure~9.
The evolution of the {\Lya} lines with
$W_r({\Lya}) > 0.3$~{\AA} can be characterized by a
double power law with $\gamma \sim 2$ for 
$1.8 < z < 4.5$ and $\gamma \sim 0.2$ for $z<1.8$.
Help in understanding the physical picture has come from
sophisticated N--body/hydrodynamic simulations that
incorporate the gas physics and consider cosmological
expansion of the simulation box.
The dynamical evolution of the {\HI} gas can
be described as outflow from the centers of
voids to their surrounding shells, and flows along
these sheets toward their intersections where
the densest structures form.
This picture is consistent with observational
determinations of the ``sizes'' of {\Lya}
structures.
It is difficult to obtain direct measurements of
sizes except in some special cases to use 
``double lines of sight'', close quasar pairs, either 
physical or apparent due to gravitational lensing.
If the spectra of the two quasars both have a {\Lya}
absorption line at the same wavelength that implies a
``structure'' which covers both lines of sight.
From these studies, it is found that
``structures'' are at least hundreds of kpc in extent.

At redshifts $z=5$ to $z=2$ $dN/dz$ for {\Lya} forest 
absorption is quite large, but it is declining
very rapidly over that range.  This dramatic evolution in the
number of forest clouds is mostly due to the
expansion of the universe, with a modest contribution
from structure growth. 
At $z<2$, the extragalactic background radiation field
is falling, and {\Lya} structures are becoming more neutral.
Therefore, the more numerous, smaller $N(H)$ structures are
observed at a larger $N({\HI})$ and this will
counteract the effect of expansion, thus slowing
the decline of the forest.

The high redshift {\Lya} forest was once thought to
be primordial material, but in fact it is observed
to have a metallicity of $0.1$\% solar, even at $z=3$.
For $N({\HI}) < 10^{14}$~{\cmsq}, the expected $N({\CIV})$ would
be below the detection thresholds of current observations, 
so truly pristine material still eludes us.
Perhaps it does not exist.
To spread metals all through the intergalactic medium
may have required a ``pre--galactic'' population of stars
at $z>10$ that polluted all of intergalactic space.

\subsection{Lyman Limit and Metal Line Systems}

The $dN/dz$ of Lyman limit systems is consistent with that
of strong {\MgII} absorbers [with $W_r({\MgII}) > 0.3$~{\AA}]
over the redshift range for which both have been observed,
$0.4 < z < 2.2$.
For $W_r({\MgII}) > 0.3$~{\AA}, $\gamma = 1.0 \pm 0.1$, consistent 
with no evolution.
For even stronger {\MgII} systems [$W_r({\MgII}) > 1$~{\AA}], 
$dN/dz$ increases more dramatically with $z$, with
$\gamma = 2.3\pm1.0$.

The number of {\MgII} systems (equivalent width distribution)
continues to increase down to the
sensitivity of the best surveys, $W_r({\MgII}) > 0.02$~{\AA},
such that $dN/dz = 2.7\pm0.15$ at $z\sim1$.
The ``weak'' {\MgII} absorbers are therefore more common than 
the strong systems [$W_r({\MgII}) > 0.3$~{\AA}] known to be 
associated with luminous galaxies.
Unlike the strong {\MgII} absorbers, the weak {\MgII} absorbers
are sub-Lyman limit systems (they do not have Lyman limit
breaks), and no galaxies have been identified at the redshift
of absorption.
Yet, photoionization models indicate that the metallicities
of these weak absorbers are at least $10$\% of the solar
value, and in some cases comparable to solar.
They are a varied population: some have relatively strong
{\FeII} while others have no {\FeII} detected,
and some have strong {\CIV} that requires a separate phase
while others have no {\CIV} detected.
Those with strong {\FeII} are constrained to be smaller than 
10~pc (the ionization parameter must be small and $n_e$ large
as can be seen in Figure~6).  Also, since Fe is produced primarily by 
Type Ia supernovae they must be enriched by a relatively old 
stellar population.  Those with weaker or undetected {\FeII} 
could be larger (kpcs or tens of kpcs) and possibly enriched by 
Type II supernovae.
Candidate environments that could be traced by weak
{\MgII} absorption are: remnants of pre--galactic star clusters 
formed in mini--halos at $z>10$, super star clusters formed in
interactions, tidally stripped material, low surface brightness
galaxies, and ejected or infalling clouds (analogous to the Milky
Way high velocity clouds).

The evolution of $dN/dz$ for {\CIV} absorbers can be studied
in the optical for high redshifts.
For $W({\CIV}) >0.4$~{\AA} and $z>1.2$, the number decreases 
with increasing $z$, as $\gamma = -2.4\pm0.8$.
In this same interval, the number of Lyman limit systems is
still increasing with redshift, with $\gamma = 1.5\pm0.4$.
This implies that the dramatic evolution in the number of
{\CIV} systems is either due to a change in metallicity
or a change in ionization state.
The $dN/dz$ for {\CIV} systems peaks at intermediate $z$
and declines, consistent with no evolution until the present.
Combining optical and UV data, {\CIV} and {\MgII} have been
compared at $0.4 < z < 2.2$.
The fraction of systems with large $W_r({\CIV})/W_r({\MgII})$ 
decreases rapidly with decreasing redshift; there is a shift
toward ``lower ionization systems''.

It is important to consider that the {\HI},
{\MgII}, and {\CIV} absorption do not always arise in the 
same phase.
It is possible that the {\CIV} in many $z\sim1$ {\MgII}
absorption systems arises in a phase similar to the
Galactic coronae.
If the origin of this phase is related to star--forming
processes in the disk, then it might be expected to
diminish below $z=1.2$ since the peak star formation rate 
is passed.

Another important trend is the fact that the very strongest
{\MgII} absorbers evolve away from $z=2$ until the present.
If we study the kinematic structure of these objects, we
find that they commonly have a ``double'' structure,
with two separate kinematic regions in the {\MgII} profile.
These objects also have strong {\CIV} which also has separate
components around the two {\MgII} regions in the ``double''
structure.  The {\CIV} does not arise primarily in the 
individual {\MgII} clouds, nor is it in a smooth, 
``common halo'' structure that extends in velocity space 
around the entire {\MgII} profile.
As more data are collected on the kinematic structure of
various transitions in these ``double'' systems, it will be 
interesting to consider the hypothesis that galaxy pairs in 
the process of merger are responsible.
The number of these is thought to have been dramatically larger
in the past.

\subsection{Damped {\Lya} Systems}

The $N({\HI}) > 10^{20.3}$~{\cmsq} systems are of particular interest
because it is possible to observe many different chemical
elements (such as Zn, Cr, Fe, Mn, and Ni) in these objects 
back to high redshift.
Metallicities and abundance patterns can be studied and
compared to those of old stellar populations in the Milky Way.
Back to $z=3$, the metallicity in DLAs, as measured by the 
undepleted element Zinc, is about 10\% 
of the solar value, but it may decline at $z>3$.
The identity of sites responsible for DLAs at high $z$ remains 
controversial, but they do contain most of the neutral Hydrogen
in the universe, from which most of its stars form.
The kinematic structure of the absorption profiles of neutral
and low ionization species is consistent with the rotation
of a thick disk, so that it is possible that these are the
$z=3$ progenitors of normal spiral galaxies.
However, this signature is not unique.
It could also be the consequence of directed infall in an
hierarchical structure formation scenario.
The higher ionization species show complex kinematics which
vary in relation to those of the lower ionization gas; in some
systems they appear to trace relatively similar structure, and
in others there are clearly several different phases.

At low redshift, many of the galaxies that are 
responsible for the DLA absorption can be directly identified.
These galaxies are a heterogeneous population.
They are not just the most luminous galaxies, but include
dwarf and low surface brightness galaxies, and even cases
where no galaxy has been identified to sensitive limits.
Damped {\Lya} absorption does not trace the most luminous
objects, but rather it traces the largest neutral gas reservoirs.
An additional selection effect may be important.
The most dust--rich galaxies that have the potential to
produce DLA absorption could produce enough extinction that
their background quasars will not be included in quasar surveys.
In this way, the population of DLAs that are actually observed
could be significantly biased against dusty galaxy hosts.

\section{Future Prospects}

The next decade will see the synthesis of the various techniques
for the study of galaxy evolution, through their stars and
through their gas.
Higher resolution quasar spectra will be obtained in the ultraviolet 
(with the Space Telescope Imaging Spectrograph (STIS) and with 
the Cosmic Origins Spectrograph (COS) on the HST, and later, 
hopefully, with a larger UV space telescope).
It will then be possible to conduct a systematic analysis of the 
relationships between the different ionization species that
trace the different phases of gas in $0.4 < z < 1.5$ galaxies.
In this redshift regime, comparisons to the detailed morphological
structure and orientations of the absorbing galaxies is possible
from HST images.

Invaluable insights into the origin of quasar absorption
lines have been gleaned from absorption studies of nearby galaxies,
for which it is possible to directly observe the processes
that are involved.
Making more observations of this type will be possible 
by discoveries of bright quasars that fall behind nearby galaxies.
The discoveries of quasars in large surveys will also
include multiple lines of sight behind distant 
absorption line systems which can be used to produce 3--D
maps of the structures.

The interstellar medium of the Milky Way shows structure on
sub--pc scales, and absorption features can only be resolved
with resolution $<1$~{\kms}.
Such a resolution will soon be available on 8m--class telescopes.
This is important for separating blends and for looking for
metallicity, ionization, and abundance pattern gradients
along the line of sight.

The key low ionization transitions of {\MgII} and {\FeII} are
shifted into the near--IR region of the spectrum for $z>2.5$.
Very soon, near--IR quasar spectra will be obtained at 
relatively high resolution ($\sim 20$~{\kms}).
Also, IR--imaging, narrow--band techniques, and multi--object
spectroscopy in the near--IR should provide much more
information about absorbing galaxies at higher redshifts.
This will extend evolutionary studies back to an epoch at
which formation processes may be contributing significantly
to evolution.

\section{Bibliography}

\noindent
Articles in review journals:
\smallskip

\noindent
Rauch M 1998 The Lyman Alpha Forest in the Spectra
of QSOs {\it ARAA} {\bf 36} 267
\smallskip

\noindent
Churchill C W and Charlton J C 2000 {\MgII} Absorbers:
A Review {\it PASP} in press
\medskip

\noindent
Conference proceedings:
\smallskip

\noindent
Blades J C, Turnshek D A, and Norman C 1988 {\it QSO Absorption
Lines: Probing the Universe, Proceedings of the QSO Absorption
Line Meeting, Baltimore 1987} (Cambridge: Cambridge University
Press)
\smallskip

\noindent
Meylan G 1995 {\it QSO Absorption Lines: Proceedings of the
ESO Workshop, Munich 1994} (Berlin: Springer)
\smallskip

\noindent
Petitjean P and Charlot S 1997 {\it Structure and Evolution of
the Intergalactic Medium from QSO Absorption Lines}
(Paris: Editions Fronti\'eres)

\end{document}